\documentclass[conference]{IEEEtran}
\IEEEoverridecommandlockouts
\usepackage{cite}
\usepackage{amsmath,amssymb,amsfonts}
\usepackage{algorithmic}
\usepackage{graphicx}
\usepackage{textcomp}
\usepackage{comment}
\usepackage{makecell}
\usepackage{multirow}
\usepackage{placeins}
\usepackage{float}
\usepackage{amssymb}
\usepackage{hyperref}
\usepackage{booktabs} 
\usepackage{tablefootnote}
\usepackage{subcaption}
\usepackage{setspace}

\def\BibTeX{{\rm B\kern-.05em{\sc i\kern-.025em b}\kern-.08em
    T\kern-.1667em\lower.7ex\hbox{E}\kern-.125emX}}
\begin{document}

\title{Post-training for Deepfake Speech Detection}
 \author{
     \IEEEauthorblockN{Wanying Ge, Xin Wang, Xuechen Liu, Junichi Yamagishi}
     \IEEEauthorblockA{
         \textit{National Institute of Informatics}, Tokyo, Japan \\
         \{gewanying, wangxin, xuecliu, jyamagis\}@nii.ac.jp
     }
 }

\maketitle

\begin{abstract}
We introduce a post-training approach that adapts self-supervised learning (SSL) models for deepfake speech detection by bridging the gap between general pre-training and domain-specific fine-tuning. We present AntiDeepfake models, a series of post-trained models developed using a large-scale multilingual speech dataset containing over 56,000 hours of genuine speech and 18,000 hours of speech with various artifacts in over one hundred languages. Experimental results show that the post-trained models already exhibit strong robustness and generalization to unseen deepfake speech. When they are further fine-tuned on the Deepfake-Eval-2024 dataset, these models consistently surpass existing state-of-the-art detectors that do not leverage post-training. Model checkpoints\footnote{Zenodo: \url{https://doi.org/10.5281/zenodo.15580542}\newline\hspace*{1.5em}Hugging Face: \url{https://huggingface.co/nii-yamagishilab}} and source code\footnote{GitHub: \url{https://github.com/nii-yamagishilab/AntiDeepfake}} are available online.

\end{abstract}


\section{Introduction}

Self-supervised learning (SSL) has significantly advanced the development and performance of deepfake speech countermeasures (CMs)~\cite{tak22_odyssey,huang2025_speechfake,liu2025nes2net}. In deepfake detection, the SSL models are used as feature extraction modules and are normally pre-trained on thousands of hours of speech data with self-supervised objectives to make them capable of producing generalized, powerful latent representations that have proven effective in solving downstream tasks~\cite{baevski2020_wav2vec,hsu2021_hubert, SUPERB, wang2018-glue,zaiem23b_interspeech}. 

However, the self-supervised objective is different from that of deepfake detection where the primary objective is to discriminate genuine speech (i.e., speech uttered by real human speakers) from speech with artifacts. Therefore, the representations extracted from pre-trained SSL models are normally further optimized using domain specific data during the fine-tuning phase every time for each deepfake dataset\cite{wang22_odyssey, tak22_odyssey, liu2025nes2net,combei24_asvspoof}. 
However, learning such meaningful representations for each of the many deepfake datasets is resource consuming. There are also various new deepfake detection tasks, such as partial spoof~\cite{zhang2022partialspoof,luong2025llamapartialspoof}  and source tracing~\cite{klein24_interspeech,xie24_interspeech}, so it is inefficient to learn meaningful representations for each of these tasks separately. \textit{Is it possible to create a new foundation model from the existing SSL model that extracts discriminative representations better suited to deepfake detection?}

\begin{figure}[t]
\centering
\includegraphics[trim=170 460 410 130, clip,  width=1\columnwidth]{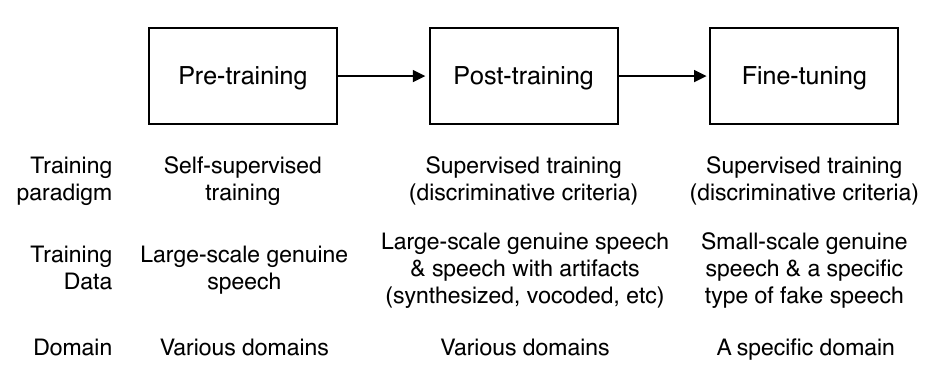}
\caption{The proposed post-training process for deepfake related tasks.}
\vspace{-0.6cm}
\label{fig:post-training}
\end{figure}

To achieve this goal, we adopt \textbf{post-training}, an important step between initial self-supervised learning and fine-tuning for the final optimization\cite{xu-etal-2019-bert,tie2025survey,qwen2-5}. This additional training phase helps us update the SSL model, making it easier to obtain useful representations for various deepfake datasets and related tasks.
While both pre-training and post-training use huge amounts of speech data, they differ significantly in data type and objective (Fig.~\ref{fig:post-training}). The proposed post-training uses genuine speech and, more importantly, various types of speech with artifacts (synthesized speech, converted speech, vocoded speech, restored speech, codec speech etc). We also introduce a  discriminative objective so that representations are more suitable for detecting speech with artifacts. While both post-training and fine-tuning aim to adapt models to a specific task, they differ significantly in scope and purpose. Post-training is performed on large, diverse datasets and aims to equip the model with domain-irrelevant representations~\cite{tie2025survey}. Fine-tuning, on the other hand, is performed on a narrow small target dataset and focuses solely on optimizing performance for that specific domain.

In this work, we developed AntiDeepfake models, a series of foundation models post-trained for deepfake detection, using a large-scale speech dataset comprising over 56,000 hours of genuine speech and 18,000 hours of speech with various artifacts in over one hundred languages.  We applied the post training using the 74k hours of speech data to both wav2vec 2.0-based~\cite{baevski2020_wav2vec,pratap2024_mms,babu22_xlsr} and HuBERT-based~\cite{hsu2021_hubert} SSL models with various sizes. We conducted two types of experiments and we found that post-trained models already exhibited strong robustness and generalization to unseen deepfake speech without fine-tuning. We also found that when they were further fine-tuned on the Deepfake-Eval-2024 dataset~\cite{chandra2025deepfake}, these models consistently surpassed existing state-of-the-art detectors that do not leverage post-training.

\section{Related work}
\subsection{Speech SSL models}
\label{sec:models}

Popular speech SSL models include wav2vec 2.0~\cite{baevski2020_wav2vec,pratap2024_mms,babu22_xlsr} and HuBERT~\cite{hsu2021_hubert}. 
The wav2vec 2.0 architecture utilizes a multi-layer convolutional encoder to process raw speech waveforms into deep feature representations and a Transformer~\cite{NIPS2017_attention} to capture contextual information from the entire sequence. The model is pre-trained in a self-supervised manner by predicting latent speech representations from randomly masked inputs. HuBERT also uses contextual information from the entire sequence and predicts cluster indexes derived from $k$-means of speech features corresponding to masked speech regions. However, these self-supervised objectives are not designed to distinguish genuine speech from deepfake speech and are therefore fundamentally different from those of deepfake detection.

\subsection{SSL model training using vocoded speech}

Previous attempts have been made to train SSL models for deepfake detection. For instance, \cite{10446331} proposed using SSL models trained on several types of vocoded speech and using the difference between the original SSL representations and ones trained on vocoded speech for deepfake detection. While the effectiveness of this method was confirmed, the use of two SSL models for feature extraction or distilling the difference between the two SSL models into a student model was either inefficient or complicated~\cite{10446331}.

\section{Post training for deepfake detection}

\subsection{Training set}
\label{sec:training_set}

Post-training is a supervised learning phase designed to adapt SSL models for deepfake detection by exposing them to diverse artifact types. To this end, we curated a diverse training set for post-training. As summarized in Table~\ref{tab:data}, our training set combines speech files from 27 publicly available sources and two self-generated datasets. These sources are categorized on the basis of artifact type:

\begin{table}[t!]
\caption{Statistics of datasets used for post-training SSL models (upper section) and their performance evaluation (bottom section). Datasets are marked as multilingual if the exact number of languages is unknown.}
\centering
\renewcommand{\arraystretch}{1.0}
\resizebox{\columnwidth}{!}{
\begin{tabular}{lcrrc}
\toprule
Dataset & Language & Genuine (hrs) & Fake (hrs) & Attack \\ \midrule 
\textit{\textbf{TTS and VC}} \\
ASVspoof2019-LA~\cite{wang2020_asvspoof2019} & en  & 11.85   & 97.80    & TTS, VC \\\midrule
ASVspoof2021-LA~\cite{yamagishi2021_asvspoof2021} & en &  16.40    & 116.10    & TTS, VC \\\midrule
ASVspoof2021-DF~\cite{yamagishi2021_asvspoof2021} & en &  20.73    & 487.00    & TTS, VC \\\midrule
ASVspoof5~\cite{wang2024_asvspoof5} & en &  413.49    & 1808.48    & TTS, VC \\\midrule
CFAD~\cite{ma2024_cfad} & zh &  171.25    & 224.55   & TTS \\\midrule
DECRO~\cite{ba2023_decro} & en, zh  &  35.18    & 102.44   & TTS, VC \\\midrule
DFADD~\cite{du2024_dfadd} & en  &  41.62    & 66.01   & TTS \\\midrule
Diffuse or Confuse~\cite{firc2024_diffuse} & en  &  0    & 231.66   & TTS \\\midrule
DiffSSD~\cite{bhagtani2024_diffssd} & en  &  0    & 139.73   & TTS \\\midrule
DSD~\cite{doan2024_dsd} & en, ja, ko  &  100.98    & 60.23   & TTS, VC \\\midrule
HABLA~\cite{florez2023_habla} & es  &  35.56    & 87.83   & TTS, TTS-VC \\\midrule
MLAAD~\cite{muler2024_mlaad} & 38 languages  &  0    & 377.96   & TTS \\\midrule
SpoofCeleb~\cite{jung2025_spoofceleb} & Multilingual  &  173.00    & 1916.2   & TTS \\\midrule
VoiceMOS~\cite{huang22f_voicemos} & en  &  0    & 448.44   & TTS \\\midrule

\textit{\textbf{Vocoded speech}}\\
CVoiceFake~\cite{li2024_cvoicefake} & en, fr, de, it, zh  &  315.14    & 1561.16   & Vocoded \\\midrule
LibriTTS~\cite{zen19_libritts} & en  &  585.83    & 0   & -- \\\midrule
LibriTTS-Vocoded & en  &  0    & 2345.14   & Vocoded \\\midrule
LJSpeech~\cite{ljspeech} & en  &  23.92    & 0   & -- \\\midrule
VoxCeleb2~\cite{chung18b_voxceleb2} & Multilingual  &  1179.62    & 0   & -- \\\midrule
VoxCeleb2-Vocoded & Multilingual  &  0    & 4721.46   & Vocoded \\\midrule
WaveFake~\cite{frank2_wavefake} & en, ja  &  0    & 198.65   & Vocoded \\\midrule

\textit{\textbf{Restored speech}} \\
FLEURS~\cite{fleurs} & 102 languages  &  1388.97    & 0   & -- \\\midrule
FLEURS-R~\cite{ma24c_fleurs-r} & 102 languages  &  0    & 1238.83   & Restored \& vocoded  \\\midrule
LibriTTS-R~\cite{koizumi23_libritts-r} & en  &  0    & 583.15   & Restored \& vocoded \\\midrule

\textit{\textbf{Neural codec speech}} \\
Codecfake~\cite{xie2025_codecfake} & en, zh &  129.66    & 808.32   & Neural codec \\\midrule
CodecFake~\cite{wu24p_codecfake} & en &  0    & 660.92   & Neural codec \\\midrule

\textit{\textbf{Additional genuine speech}} \\
AISHELL3~\cite{shi21c_aishell3} & zh  & 85.62   & ~0   & -- \\\midrule
CNCeleb2~\cite{li2022_cnceleb} & zh &  1084.34    & 0   & -- \\\midrule
MLS~\cite{pratap20_mls} & 8 languages  &  50558.11    & 0   & -- \\\midrule
\midrule
\textbf{Train Set} & Over 100 languages  &  56.37~k & 18.28~k &  -- \\ \midrule \midrule
\end{tabular}
}
\vspace{-0.55cm}
\label{tab:data}
\end{table}

\subsubsection{\textbf{Datasets with generated deepfake speech}}

We collected datasets commonly utilized in the deepfake detection community, ranging from the ASVspoof challenges~\cite{wang2020_asvspoof2019,yamagishi2021_asvspoof2021,wang2024_asvspoof5} to the recent Diverse Synthesizer for Deepfake Voice Detection Corpus (DSD-corpus)~\cite{doan2024_dsd}. These datasets offer a broad range of deepfake speech generated using text-to-speech (TTS) and voice conversion (VC) algorithms, alongside corresponding genuine speech samples.
While most deepfake datasets are dominated by English and Mandarin Chinese, we also included recent multilingual datasets such as HABLA~\cite{florez2023_habla} (Latin American Spanish), Multi-Language Audio Anti-Spoofing Dataset~\cite{muler2024_mlaad} (MLAAD, 38 languages), and SpoofCeleb~\cite{jung2025_spoofceleb} (multilingual, sourced from VoxCeleb1~\cite{nagrani17_voxceleb1})\footnote{Although it was originally designed for predicting mean opinion scores for synthetic speech, we also added the synthesized partition of the VoiceMOS dataset~\cite{huang22f_voicemos} to our training set. The human speech portion in the dataset was excluded.}.

\subsubsection{\textbf{Datasets with vocoded speech}}
\label{sec:vocoded}

To enhance the model’s ability to detect waveform-level artifacts introduced during voice generation, we included vocoded speech and their original versions as a second category. Although the speech content of these vocoded speech files remained the same as their original counterparts, their waveforms were generated using vocoders with vocoder-specific artifacts. This category includes CVoiceFake~\cite{li2024_cvoicefake} and WaveFake \cite{frank2_wavefake}, which is the vocoded version of LJSpeech~\cite{ljspeech}. We also generated extra vocoded data by processing genuine speech from LibriTTS ~\cite{zen19_libritts} and VoxCeleb2~\cite{chung18b_voxceleb2} with several neural and DSP vocoders~\cite{wang2023_copy_synthesis}. 

\subsubsection{\textbf{Datasets with restored speech}}

We included neural restored speech as a third category. Specifically, we added speech files from FLEURS-R~\cite{ma24c_fleurs-r} and LibriTTS-R~\cite{koizumi23_libritts-r} and their original low-quality versions \cite{fleurs}\cite{zen19_libritts} during post-training. The speech content and source speakers were the same as in the originals, but the audio quality was improved through a high-quality diffusion vocoder~\cite{wavefit}.  

\subsubsection{\textbf{Datasets with neural codec speech}}
\label{sec:codec}

We also included two neural codec-based datasets as a fourth category of our post-training set: Codecfake~\cite{xie2025_codecfake} and CodecFake~\cite{wu24p_codecfake}. Both datasets contain speech files regenerated by applying recent neural codec algorithms to reconstruct genuine speech. Note that codec speech is not to be regarded as deepfake speech, but as one of the post-training tasks of speech with artifacts, and the goal of the post training was to obtain a representation suitable for detecting fine differences.

\subsubsection{\textbf{Datasets with additional genuine speech}}
\label{sec:genuine_data}

To further diversify the training set, we incorporated genuine-only multilingual datasets covering various speakers, languages, and acoustic conditions. Specifically, we included AISHELL3~\cite{shi21c_aishell3}, CNCeleb2~\cite{li2022_cnceleb}, each of which is commonly utilized in speaker recognition and offers a wide range of speakers and recording conditions, and Multilingual LibriSpeech (MLS)~\cite{pratap20_mls}.

\subsection{Data pre-processing and post-training process}
\label{sec:data_partition}
\begin{figure}[t]
\centering
\includegraphics[trim=0 110 0 145, clip,  width=1\columnwidth]{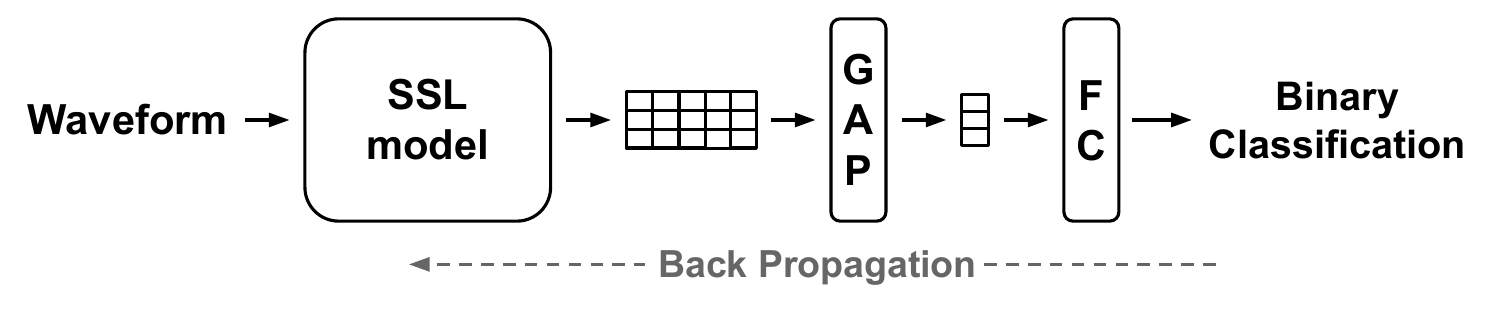}
\caption{Model architecture used for post-training. It comprises a self-supervised front-end, a global average pooling layer (GAP), and a fully connected layer (FC).}
\vspace{-0.3cm}
\label{fig:cm}
\end{figure}

All speech files used in this study were resampled to 16 kHz, converted to a single channel, and normalized. 
All post-training reported in this paper was in the same manner as shown in Fig~\ref{fig:cm}. A pre-trained SSL model was used as an initial weight for extracting sequence-level representations from the input speech data. These representations were directly fed into a global average pooling layer, which aggregated the temporal information, followed by a fully connected (FC) layer that generated an output score indicating the likelihood of the input utterance being genuine speech without artifacts belonging to one of categories above. Post-training was then performed using backpropagation based on a cross-entropy loss function and updating the weights of the SSL model. Here, we compute the sums of the cross entropy of each category without any weighting. Both the SSL model and FC layer were updated jointly during the post-training stage.
The model after post-training can be used to extract feature for deepfake detection in a zero-shot manner, or it may be further fine-tuned and optimized for a specific dataset or task.

\section{Experiments}
\label{sec:results}
\begin{table*}[h]
\renewcommand{\arraystretch}{1.02}
    \centering
    \caption{
    Equal Error Rate (EER) results of different models on various test datasets under zero-shot evaluation, without any fine-tuning for all systems. For post-training, each cell presents results in the format: “with RawBoost~/~no RawBoost”. Values in \textbf{bold} indicate the best results among post-trained SSL models, while \underline{underlined} values highlight the best among baseline systems for each test dataset. No publicly available EER results have been reported for DEEP-VOICE as an unseen test dataset.
    }

\vspace{-0.15cm}
    \resizebox{0.96\textwidth}{!}{
    \begin{tabular}{c r r r r r r r r}
        \toprule
        & \multirow{2}{*}{Model ID} & \multirow{2}{*}{\# of params.} & ADD 2023 & DEEP-VOICE & \multicolumn{2}{c}{FakeOrReal} & In-the-Wild & Deepfake-Eval-2024 \\
        \cmidrule(lr){4-4} \cmidrule(lr){5-5} \cmidrule(lr){6-7} \cmidrule(lr){8-8} \cmidrule(lr){9-9}
        && & Track-1.2-R2-Test & Segmented Full Set & original-Test & norm-Test & Full Set & Audio Test \\
        \midrule
\multirow{7}{*}{\rotatebox{90}{\shortstack{Pre-training\\+ Post-training}}}
         & \texttt{HuBERT-XL} & 964 M & 18.91~/~35.34 & 5.68~~/~14.84 & 2.48~~/~~3.67 & 3.18~~/~15.56 &5.23~~/~17.99 & 34.10~/~47.73\\
         & \texttt{W2V-Small} & 95 M & 13.02~/~19.41 & 9.79~~/~16.20 & 22.03~/~~1.06 & 17.89~/~~6.48 & 4.23~~/~~4.65 & 33.33~/~31.97\\
         & \texttt{W2V-Large} & 317 M & 13.25~/~12.67 & 4.44~~/~~5.04 & 0.67~~/~~0.80 & \textbf{0.97}~~/~~1.45 & 1.91~~/~~2.25 & 33.36~/~30.06\\
         & \texttt{MMS-300M} & 317 M & 7.93~~/~11.22 & 2.35~~/~~3.06 & 1.40~~/~~\textbf{0.48} & 5.92~~/~~2.70 & 2.90~~/~~2.00 & 32.84~/~31.39\\
         & \texttt{MMS-1B} & 965 M & 9.06~~/~~9.46 & 2.47~~/~~2.35 & 1.23~~/~~0.88 & 1.73~~/~~1.10 & 1.82~~/~~1.87 & 27.70~/~27.54\\
         & \texttt{XLS-R-1B} & 965 M & 5.39~~/~~6.58 & 2.47~~/~~2.59 & 5.74~~/~~3.17 & 12.14~/~10.91 & 1.35~~/~~1.36 & 26.76~/~26.17\\
         & \texttt{XLS-R-2B} & 2.2 B & \textbf{4.67}~~/~~6.84 & \textbf{2.23}~~/~~2.72 & 2.61~~/~~1.19 & 1.64~~/~~1.73 & \textbf{1.23}~~/~~1.31 & 27.76~/~\textbf{25.76}\\
\midrule
        \multirow{11}{*}{\rotatebox{90}{\shortstack{Zero-shot evaluation \\results in the literature}}}& XLSR-Mamba~\cite{xiao2024xlsr-mamba} & 319 M & 19.36 & - & 6.71 & - & 6.70 & - \\
        & Resemble AI~\cite{speecharena-df-leaderboard} & 2.1 B & \underline{6.11} & - & 1.36 & - & 3.94 & -  \\
        & SpeechFake~\cite{huang2025_speechfake} & 317 M & - & - & 4.88 & - & 2.01 & - \\
        & Wav2Vec + VIB~\cite{doan2024_dsd} & - & - & - & - & \underline{3.93} & \underline{1.99} & - \\
        & UniSpeech-SAT~\cite{unispeech-sat,speecharena-df-leaderboard} & 96 M & 28.21 & - & \underline{1.06} & - & 15.05 & -\\
        & XLS-R + SLS~\cite{xlsr-sls} & 340 M & 21.10 & - & 5.08 & - & 7.45 & - \\
        & XLSR-Conformer + TCM~\cite{truong24b_interspeech} & 319 M & 22.74 & - & 10.69 & - & 7.79 & - \\
        & AdaLAM \& f-InfoED~\cite{zhao2025generalized} & - & - & - & - & - & 8.36 & -\\
        & P3~\cite{wang2024can,chandra2025deepfake} & 317 M & - & - & - & - & - & \underline{43.00} \\ 
        & AASIST~\cite{jung2022aasist,chandra2025deepfake} & 0.3 M & 32.47 & - & 21.64 & - & 43.01 & 55.22 \\
        & RawNet2~\cite{Tak2021rawnet2,chandra2025deepfake} & 18 M & 64.55 & - & 65.68 & - & 49.19 & 48.20\\
        \bottomrule
    \end{tabular}
    }
    \label{tab:results}
    \vspace{-0.25cm}

\end{table*}

\subsection{Experimental conditions}
\label{sec:setup}

We used wav2vec 2.0 and HuBERT of various sizes pre-trained on a variety of data shown in Table~\ref{tab:results} for post training and evaluation. Speech files originally designated for validation in their respective datasets are used for best model selection, and the remaining files are used for model learning. 
The training data was sampled dynamically by grouping files of similar durations and zero-padding them to form mini-batches~\cite{speechbrainV1}. Files longer than 13 seconds were randomly trimmed to durations between 10 and 13 seconds during training. We also used RawBoost with its best configurations from~\cite{tak2022_rawboost} as a data augmentation method. For the testing, all files were kept at their original durations without augmentation. 

The maximum batch duration is set to 100 seconds for all models except the largest one, \texttt{XLS-R-2B}, which has a maximum batch duration of 50 seconds. We performed validations every 100k mini-batches (steps). 
All the experiments used an AdamW optimizer~\cite{loshchilovDecoupled2019} with the value of weight decay set to 0.01.
Except for \texttt{HuBERT-XL} (which uses a learning rate of 5e-6), all models were optimized with the learning rate linearly increasing to 1e-7 over the first 80k steps and then linearly decaying in steps of 800k to 0, at which point the training stopped. This allowed \texttt{XLS-R-2B} to be updated with over 90\% of the training data at least once, while other models saw most of the training data twice. All experiments were conducted with the same random seed on eight NVIDIA H100 GPUs.

\subsection{Multiple test sets}
\label{sec:test_set}
Our test set included FakeOrReal~\cite{for} and its second version FakeOrReal-norm, In-the-Wild~\cite{muller22_inthewild}, DEEP-VOICE~\cite{bird2023_deepvoice}, the Audio Deepfake Detection Challenge (ADD 2023) track 1.2 evaluation set~\cite{yi2023_add2023} and the Deepfake-Eval-2024 dataset~\cite{chandra2025deepfake}. None of them were used for pre-training and post-training.  

FakeOrReal and In-the-Wild are English-based datasets that are widely used in the deepfake detection community to assess a CM’s generalization performance. ADD 2023 is a Mandarin-based challenge dataset, and its deepfake speech samples are collected from the generation task of the same challenge. DEEP-VOICE consists of genuine presentation recordings from eight celebrities, along with their converted versions. Lastly, the Deepfake-Eval-2024 dataset contains diverse genuine and deepfake multimedia content collected in 2024 from social media and deepfake detection platforms. In our study, we used only its audio portion, which includes samples in more than 50 languages.

We conducted two types of experiments, zero-shot and fine-tuned performance evaluations. 
For the zero-shot performance evaluation described in Sec.~\ref{sec:exp:zeroshot}, we used only the test partitions of the selected test datasets; their training portions, if available, were excluded entirely from both the model training and testing. Note that the original utterances in DEEP-VOICE and Deepfake-Eval-2024 are unsegmented and can last several minutes, which requires a huge amount of GPU memory for not only the SSL-based but also other deep neural network-based deepfake CMs. We therefore manually divided them into shorter segments before adding them to our test set. For DEEP-VOICE, our released code provides the timestamps used for segmentation in our released code. For Deepfake-Eval-2024, we used the approach in~\cite{chandra2025deepfake} to divide each audio file in the test set into non-overlapping 4-second segments, used the CMs to score each segment independently, and measured the equal error rates (EERs) over the segments. 
We also conducted the evaluation multiple times, wherein the audio file was divided into segments of 10, 13, 30, and 50 seconds, respectively. The 13-second duration was included because it corresponds to the maximum input length used during both post-training and fine-tuning. The training files from the Deepfake-Eval-2024 were of the its original length for the fine-tuning stage described in Sec.~\ref{sec:finetuning}.

\subsection{Zero-shot evaluation results}
\label{sec:exp:zeroshot}

First, we analyzed the results of zero-shot evaluation results. The post-trained models were directly used for the test sets without any fine-tuning process. Table~\ref{tab:results} presents the results of the EER evaluation results of the models post-trained with and without data augmentation. Models are grouped by architecture (HuBERT and wav2vec 2.0) and sorted by their number of parameters.

We can see that although the post-trained models were not fine-tuned to any of these test sets, they exhibited strong robustness and generalization to unseen deepfake speech.
The results generally improved as the model size increased, apart from \texttt{HuBERT-XL}. The best EER results for ADD 2023 (4.67\%), DEEP-VOICE (2.23\%), and In-the-Wild (1.23\%) were achieved by our largest model \texttt{XLS-R-2B}. On the other hand, on both FakeOrReal test sets, the smaller models tended to perform better, with the best results for these sets coming from SSL models with 317 M parameters (0.48\% for \texttt{MMS-300M} and 0.97\% for \texttt{W2V-Large}). Notably, \texttt{W2V-Large} achieved the best result for the FakeOrReal-norm (0.97\%) and the second-best for the original FakeOrReal (0.67\%). Additionally, all models performed poorly on the Deepfake-Eval-2024 dataset, though larger models still tended to offer slightly better performance.

Regarding the use of RawBoost-based data augmentation, the results in Table~\ref{tab:results} do not indicate that it is always beneficial. In fact, many CMs performed worse on the FakeOrReal datasets when they were trained with RawBoost. However, its use was almost always beneficial for the models tested on the ADD 2023, DEEP-VOICE, and In-the-Wild datasets. 

\subsection{Fine-tuning and evaluation on Deepfake-Eval-2024}
\label{sec:finetuning}

Among the test datasets, Deepfake-Eval-2024 is the most challenging. There are several reasons for this: it was collected from several social media platforms manually and released in 2024, making it the most recent and up-to-date; it contains a wide range of real-world recordings containing complex acoustic background conditions.
These characteristics make Deepfake-Eval-2024 more difficult than other test sets. As a result, it serves as an ideal benchmark for fine-tuning and evaluating the generalization capability of our AntiDeepfake models.

\begin{figure}[t]
\centering
\includegraphics[trim=8 0 300 0, clip,  width=0.70\columnwidth]{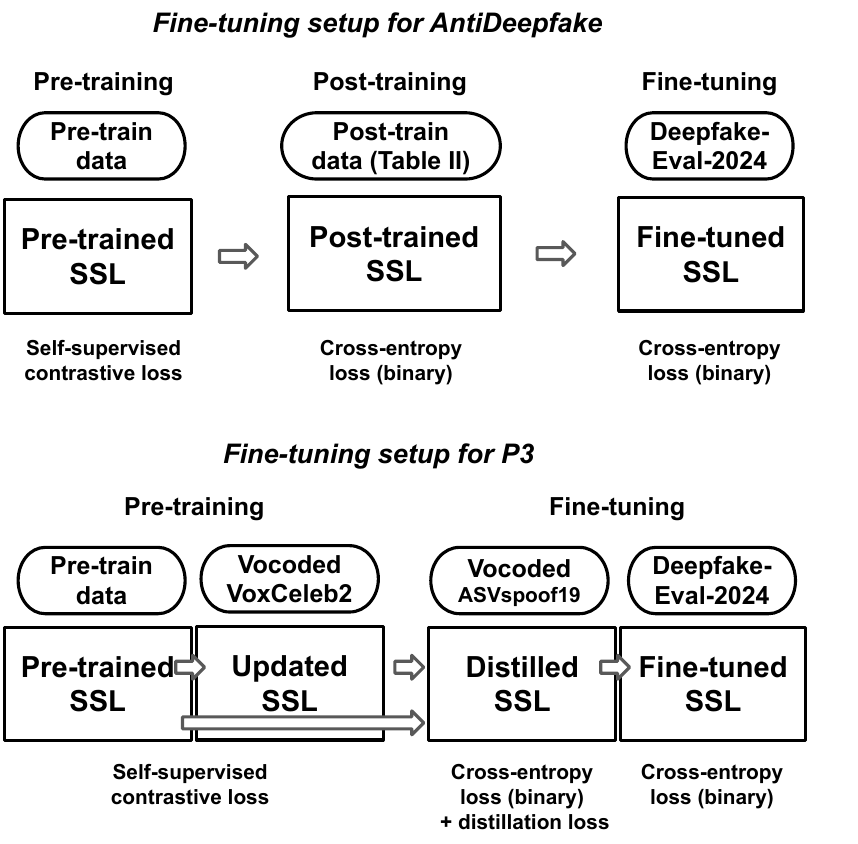}
\caption{Illustration of training setups for AntiDeepfake models (top) and the P3 model \cite{chandra2025deepfake} (bottom) for evaluation on Deepfake-Eval-2024 dataset.}
\vspace{-0.55cm}
\label{fig:exp:finetuning}
\end{figure}

For comparison, we also included an open-sourced CM called P3, which was built using a similar procedure with pre-training, another step of pre-training on large scale vocoded data, and fine-tuning~\cite{chandra2025deepfake}. Fig.~\ref{fig:exp:finetuning} illustrates the setup of these two models, and the details are explained below. 
This P3 system is highly relevant to the proposed post-training framework but uses different training criteria and multiple fine-tuning stages. The comparison is expected to give a hint on the choice of designs that could potentially improve the proposed post-training framework. Additionally, we also included non-SSL systems for reference. 

\subsubsection{\textbf{Fine-tuning setup for AntiDeepfake models}}

We fine-tuned our best-performing post-trained AntiDeepfake models (reported in Sec.\ref{sec:exp:zeroshot}) using the training portion of the Deepfake-Eval-2024 audio dataset. We randomly selected 90\% of the training files for fine-tuning and used the remaining 10\% was used for best model checkpoint selection. 
The model parameters were updated for 6k steps using the AdamW optimizer with a learning rate of 1e-6 and with RawBoost augmentation.

\subsubsection{\textbf{Fine-tuning setup for P3 model}}
The P3 model starts with a wav2vec 2.0 model pre-trained on multi-lingual human speech data (XLSR-53~\cite{xlsr53}). The pre-trained SSL model weights were updated using VoxCeleb2 and its vocoded versions, under the exact same self-supervised training criterion as in the initial pre-training stage. In the fine-tuning stage, the pre-trained SSL and the updated SSL models were distilled into a wav2vec 2.0 model of the same size, with an additional binary classification cross-entropy loss and the vocoded versions of the ASVspoof 2019 datasets~\cite{wang2020_asvspoof2019}. Global average pooling and a fully connected layer were attached to the student wav2vec 2.0 model during the fine tuning. In the last step, the fine-tuned model was further fine-tuned on the Deepfake-Eval-2024 training set. Both fine-tuning steps used the Adam optimizer with a learning rate of 1e-6 and the RawBoost augmentation. 

\subsubsection{\textbf{Fine-tuning setup for non-SSL models}}
For reference, the popular non-SSL models AASIST~\cite{jung2022aasist} and Rawnet2~\cite{Tak2021rawnet2} were included in the experiments. These two systems were first trained using the standard cross-entropy loss on the ASVspoof 2019 training set for binary deepfake detection and then fine-tuned with the same loss on the Deepfake-Eval-2024 training set. 

\subsubsection{\textbf{Evaluation results}}

Table~\ref{tab:finetuning} presents EER results for the selected AntiDeepfake models on the Deepfake-Eval-2024 test set after fine-tuning. To evaluate the effect of post-training, we compared each model's performance with and without post-training prior to fine-tuning. As aforementioned, the evaluations were conducted multiple times, wherein the input data had a duration of 4, 10, 13, 30, or 50 seconds.

First, we can see that applying post-training consistently and clearly improved EER performance across all models. We can also see that larger models tended to perform better. Among them, \texttt{XLS-R-1B} achieved the best results, with its EER decreasing from 11.86\% at 4 seconds to 8.28\% at 50 seconds—the lowest EER in the table.

We also observed that longer input durations generally lead to lower EERs, since longer segments are more likely to contain speech rather than silent regions. Notably, the benefits of post-training were most pronounced for shorter durations (4s and 10s). For example, at 4 seconds, post-training reduced the EER by 8.1\% for \texttt{XLS-R-1B} and 7.7\% for \texttt{MMS-1B}, indicating improved generalization in low-information scenarios. While 13 seconds was the maximum duration used during training, extending the input length to 30 and 50 seconds continued to reduce EERs—albeit with limited returns.

In the evaluation condition using 4-second segments, the P3 model demonstrated stronger performance than \texttt{MMS-300M} that has a similar model size. Although factors such as different SSL front ends may have affected the results, the comparison at least suggests potential techniques that may be incorporated in the proposed post-training framework. In particular, a contrastive loss may be more suitable than the cross-entropy loss for post-training. It also raises the question of whether vocoded data is sufficient for post-training. If so, the cost post-training cost can be significantly reduced since larger scale vocoded data can be created within a short time. The comparison between P3 and AntiDeepfake models provides hints for future work.

\begin{table}[t!]
\renewcommand{\arraystretch}{1.0}
    \centering
    \caption{EER results of AntiDeepfake models on the Deepfake-Eval-2024 test set across varying input durations. We compare models fine-tuned with and without prior post-training. \textbf{Bold} font indicates the best result in each column.}
    \vspace{-0.1cm}
    \resizebox{1\linewidth}{!}{
    \begin{tabular}{l ccccc ccccc}
    \toprule
    & \multicolumn{5}{c}{Pre-training + Post-training + Fine-tuning} & \multicolumn{5}{c}{Pre-training + Fine-tuning} \\
    \cmidrule(lr){2-6} \cmidrule(lr){7-11}
    Model ID & 4s & 10s & 13s & 30s & 50s & 4s & 10s & 13s & 30s & 50s \\
    \midrule
    \texttt{W2V-Large}   & 19.56 & 12.11 & 10.95 & 10.52 & 11.36 & 24.42 & 22.46 & 22.14 & 21.17 & 21.53 \\
    \texttt{MMS-300M}    & 17.15 & 13.38 & 12.32 & 11.18 & 10.79 & 19.77 & 13.26 & 12.72 & 12.01 & 12.30 \\
    \texttt{MMS-1B}      & 12.11 & 10.40 & 10.02 & 8.60  & 9.35  & 19.85 & 13.57 & 11.56 & 11.04 & 11.40 \\
    \texttt{XLS-R-1B}   & \textbf{11.86} & 10.00 & \textbf{9.27}  & \textbf{8.49}  & \textbf{8.28} & 19.96 & 17.19 & 16.31 & 10.71 & 11.20 \\
    \texttt{XLS-R-2B}    & 12.14 & \textbf{9.80}  & 9.99  & 9.46  & 9.56  & \textbf{12.88} & \textbf{10.76} & \textbf{10.34} & \textbf{9.68} & \textbf{9.84} \\
    \midrule
    P3~\cite{chandra2025deepfake} & - & - & - & - & - & 15.38 & - & - & - & - \\
    AASIST~\cite{chandra2025deepfake} & - & - & - & - & - & 16.99 & - & - & - & - \\
    RawNet2~\cite{chandra2025deepfake} & - & - & - & - & - & 20.91 & - & - & - & - \\
    \midrule
\end{tabular}
    }
\vspace{-0.4cm}
\label{tab:finetuning}
\end{table}

\subsection{Visualization of embedding representations}

\begin{figure}[t]
    \centering
    \begin{subfigure}[t]{0.73\linewidth}
        \centering
        \includegraphics[width=\linewidth]{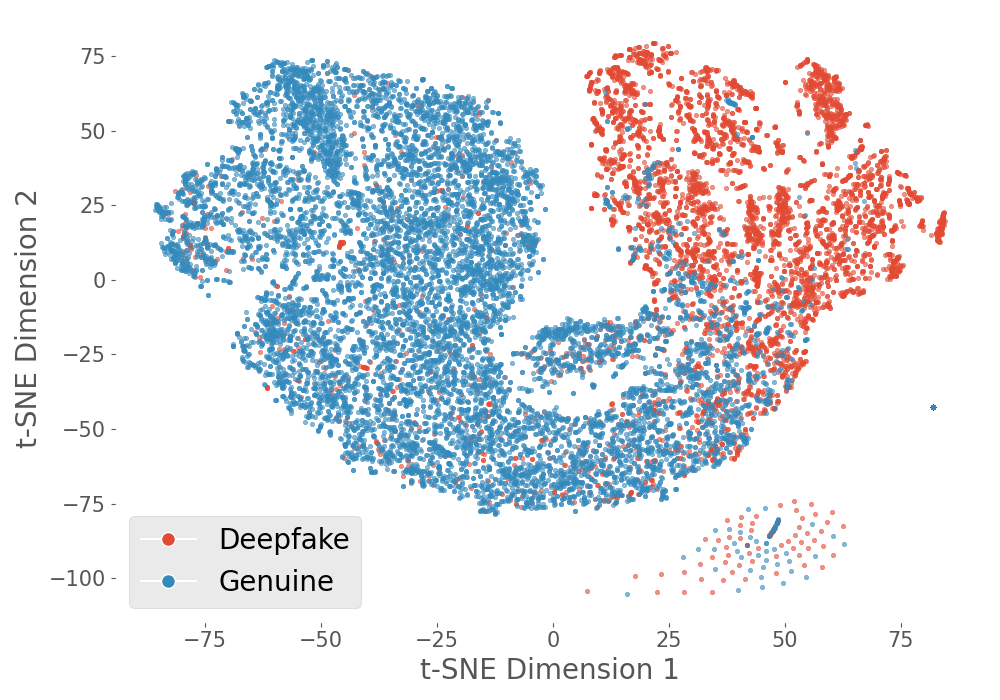}
        \caption{Pre-training + Post-training + Fine-tuning.}
        \label{fig:tsne-pt-ft}
    \end{subfigure}
    \hfill
    \begin{subfigure}[t]{0.73\linewidth}
        \includegraphics[width=\linewidth]{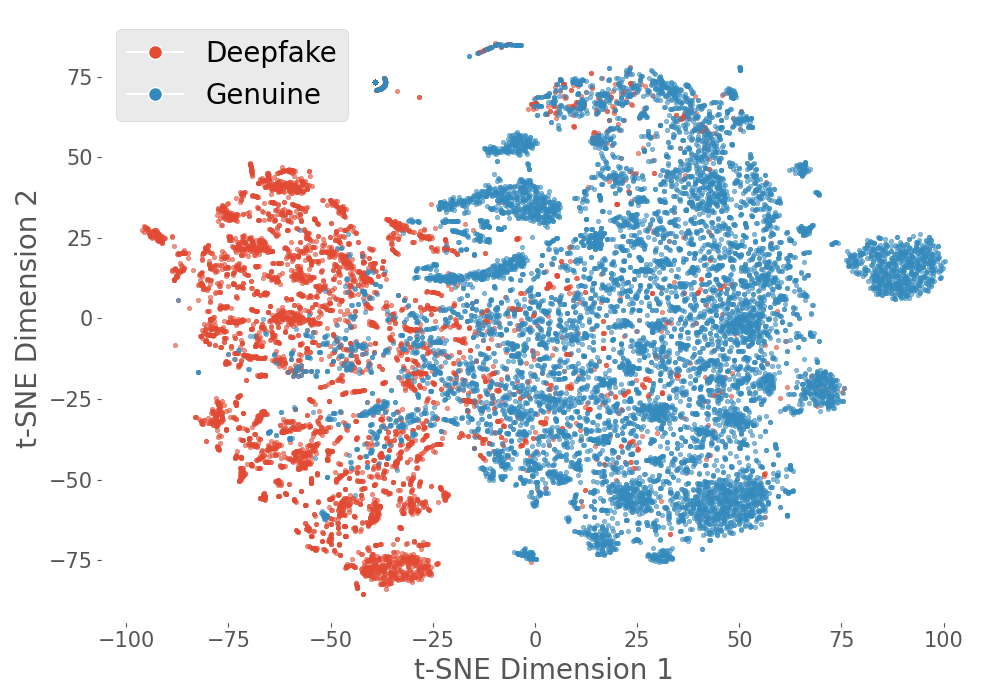}
        \caption{Pre-training + Fine-tuning.}
        \label{fig:tsne-ft}
        \centering
    \end{subfigure}
    \hfill
    \begin{subfigure}[t]{0.73\linewidth}
        \centering
        \includegraphics[width=\linewidth]{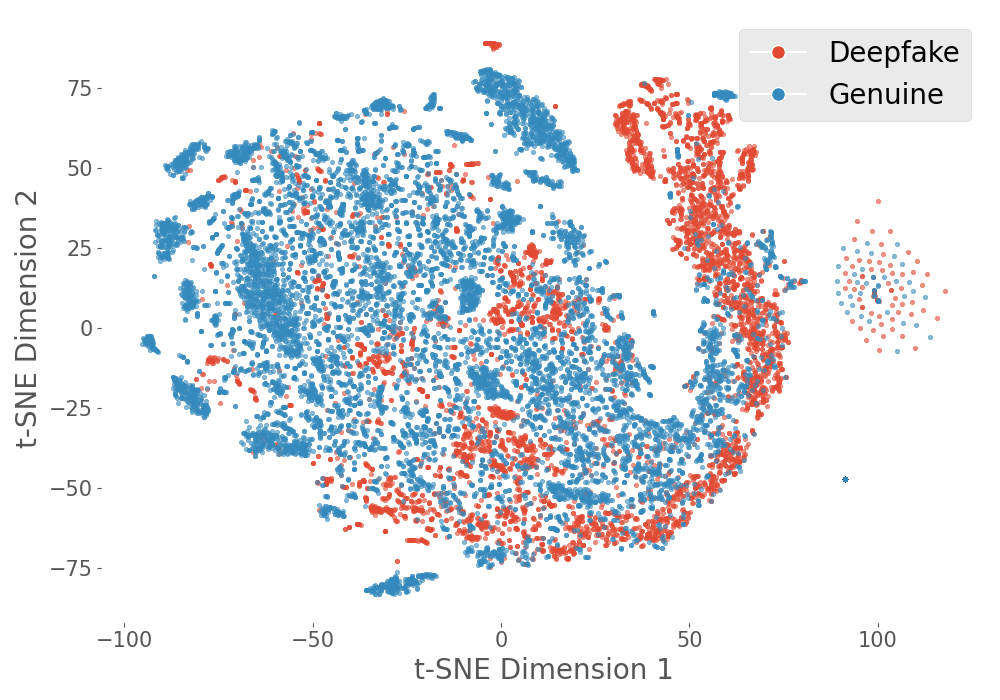}
        \caption{Pre-training + Post-training.}
        \label{fig:tsne-pt}
    \end{subfigure}
    \caption{t-SNE visualization of the embedding representation extracted by \texttt{XLS-R-2B} model after the GAP layer on the whole Deepfake-Eval-2024 test set. Blue dots stand for genuine files and red ones stand for deepfake files.}
    \vspace{-0.35cm}
    \label{fig:tsne}
\end{figure}

To analyze the effect of post-training, we used t-SNE to visualize the embeddings from our largest AntiDeepfake model, \texttt{XLS-R-2B}, as shown in Fig.~\ref{fig:tsne}. Figs.~\ref{fig:tsne-pt-ft}, \ref{fig:tsne-ft}, and \ref{fig:tsne-pt} show the embeddings of models with (a) pre-training + post-training + fine-tuning, (b) pre-training + fine-tuning, and (c) pre-training + post-training, respectively.

All three models exhibit some degree of separation between deepfake (red) and genuine (blue) samples. However, the model using only pre-training and fine-tuning (Fig.~\ref{fig:tsne-ft}) shows a more mixed and overlapping distribution compared to the fine-tuned post-trained model (Fig.~\ref{fig:tsne-pt-ft}). This visualization underpins that post-training improves the model’s ability to learn more discriminative representations.

Finally, to demonstrate the utility of the embeddings resulting from the proposed post training for other different deepfake-related tasks, we visualized part of a test set of the PartialSpoof dataset \cite{zhang2022partialspoof} containing speech in which only partial segments, rather than entire utterances, were synthesized by TTS or VC systems. Note that the test set was balanced so that the ratio of real to fake was one-to-one. Although PartialSpoof is a different task, the visualization shows that the real and fake classes are generally separated. In other words, the feature representation obtained from the post-trained model is expected to be useful for other tasks related to deepfake audio.

\begin{figure}[t]
    \centering
    \includegraphics[width=0.69\linewidth]{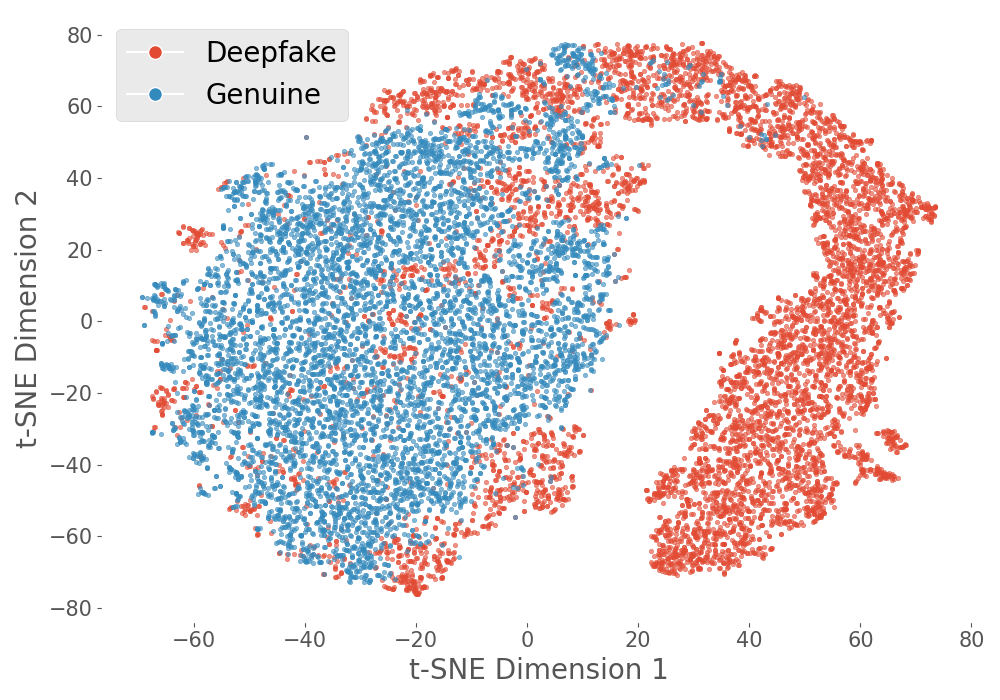}
    \label{fig:tsne-pt-partialspoof}
    \caption{t-SNE visualization of the embedding representation extracted by \texttt{XLS-R-2B} model using \textbf{pre-training and post-training}. Part of the \textbf{PartialSpoof test set} was used. Blue dots stand for genuine files and red ones for partial deepfake files.}
    \vspace{-0.3cm}
\end{figure}

\section{Conclusion and limitations}

We introduced a new post-training approach for deepfake speech detection to bridge the gap between general SSL pre-training and fine-tuning on a specific dataset for the target domain. During the post-training stage, we exposed the pre-trained SSL models to both genuine speech samples and samples containing various artifacts in various languages and domains and post-trained it to distinguish between samples with and without artifacts. The samples included not only synthesized and/or converted speech using text-to-speech and voice conversion systems but also vocoded speech, restored speech, and codec speech in over one hundred languages. Experimental results showed that this post-training enabled the model to achieve high robustness and generalization performance on unseen deep fake audio data, even when the model was used in a zero-shot manner without any fine-tuning. Furthermore, we also demonstrated that the post-trained model provides better representations for fine-tuning, and that the performance on the most challenging test set was improved by fine-tuning the post-trained model compared with just fine-tuning the pre-trained SSL models.

Although this paper demonstrated the effectiveness of post-training, some aspects require further investigation. For example, RawBoost was used as data augmentation during post-training in this study, but as was shown in the Deepfake-Eval-2024 experiment (Table~\ref{tab:results}), data augmentation using RawBoost is insufficient against the audio collected from the real world, and data augmentation using e.g. MUSAN~\cite{snyder2015_musan} to simulate complex background noise and background music under varying signal-to-noise ratios will likely be necessary. In addition, this study used the cross-entropy loss as the training criterion for post-training, but this may not be optimal. Post-training, based on other criteria such as supervised contrastive learning~\cite{SupContrast}, is also possible, as the results in Sec~\ref{sec:finetuning} indicate. Furthermore, post-training was performed without weighting the loss for multiple categories of speech artifacts, but it is also feasible to define priority categories and vary the weights during the post-training. It is also unclear whether text-to-speech and/or voice conversion-based synthetic speech, which are time-consuming to build and generate, are crucial for post-training. Vocoded, restored, and coded speech, which can be easily created from large amounts of human speech, may be sufficient for post-training. Finally, due to space limitations, we did not conduct experiments other than deepfake detection on fully spoofed audio. It is necessary to extend our investigation to source tracing and partial spoof localization using the robust features of the proposed post-trained model. These tasks are our future work.

\section*{Acknowledgment}
This paper is based on results obtained from a project, JPNP22007, commissioned by the New Energy and Industrial Technology Development Organization (NEDO). This study is partially supported by JST AIP Acceleration Research (JPMJCR24U3) and JST PRESTO (JPMJPR23P9).  This study was carried out using the TSUBAME4.0 supercomputer at the Institute of Science Tokyo.

\bibliographystyle{IEEEtran}
\bibliography{mybib}
\end{document}